\documentclass[conference]{IEEEtran}
\IEEEoverridecommandlockouts

\usepackage{cite}
\usepackage{amsmath,amssymb,amsfonts}
\usepackage{algorithmic}
\usepackage{graphicx}
\usepackage{textcomp}
\usepackage{xcolor}
\usepackage{amssymb}

\def\BibTeX{{\rm B\kern-.05em{\sc i\kern-.025em b}\kern-.08em
    T\kern-.1667em\lower.7ex\hbox{E}\kern-.125emX}}
\begin{document}

\title{Assessing Quantum Annealing to solve the Minimum vertex multicut\\
}

\author{\IEEEauthorblockN{Ali Abbassi}
\IEEEauthorblockA{\textit{Orange Research, LIST3N} \\
\textit{University of Technology of Troyes}\\
Chatillon-Troyes,  France \\
ali.abbassi@{utt.fr/orange.com}}
\and
\IEEEauthorblockN{Yann Dujardin}
\IEEEauthorblockA{\textit{Orange Research} \\
\textit{Orange}\\
Chatillon,  France \\
yann.dujardin@orange.com}
\and
\IEEEauthorblockN{Eric Gourdin}
\IEEEauthorblockA{\textit{Orange Research} \\
\textit{Orange}\\
Chatillon,  France \\
eric.gourdin@orange.com}
\and

\IEEEauthorblockN{Philippe Lacomme
 }
\IEEEauthorblockA{\textit{LIMOS - UMR CNRS 6158
ISIMA - Institut d'Informatique d'Auvergne} \\
\textit{Université Clermont Auvergne}\\
Campus Universitaire des Cézeaux
1, rue de la Chebarde
63177 Aubière, France \\
philippe.lacomme@isima.fr }
\and
\IEEEauthorblockN{Caroline Prodhon}
\IEEEauthorblockA{\textit{LIST3N} \\
\textit{University of Technoogy of Troyes}\\
Troyes, France \\
caroline.prodhon@utt.fr}
}

\maketitle

\begin{abstract}
Cybersecurity in telecommunication networks often leads to hard combinatorial optimization problems that are challenging to solve with classical methods. This work investigates the practical feasibility of using quantum annealing to address the Restricted Vertex Minimum Multicut Problem. The problem is formulated as a Quadratic Unconstrained Binary Optimization model and implemented on D-Wave’s quantum annealer. Rather than focusing on solution quality alone, we analyze key aspects of the quantum workflow including minor embedding techniques, chain length, topology constraints, chain strength selection, unembedding procedures, and postprocessing. Our results show that quantum annealing faces substantial hardware-level constraints limitations in embedding and scalability, especially for large instances, while hybrid quantum-classical solvers provide improved feasibility. This study offers a realistic assessment of the D-Wave system's current capabilities and identifies crucial parameters that govern the success of quantum optimization in cybersecurity-related network problems.
\end{abstract}


\section{Introduction}

Modern telecommunication networks must continuously contend with the challenges of ensuring secure, reliable, and efficient communication, especially in the face of adversarial attacks or system failures. Many of the algorithmic challenges that arise in this context can be framed as hard combinatorial optimization problems

A fundamental example is the \textit{Vertex Minimum Multicut (VMMC)} problem. Given an undirected graph and a set of terminal node pairs, the objective is to identify the smallest set of non-terminal vertices whose removal disconnects all specified terminal pairs. This problem arises in numerous real-world scenarios including intrusion containment, resilient routing, and the design of survivable communication infrastructure~\cite{Costa2005MulticutSurvey, Stone1977Multiprocessor}. VMMC also serves as a natural abstraction for tasks involving attack path disruption or selective isolation in security-focused network configurations.

The VMMC problem is known to be \textbf{NP-complete} even on restricted graph classes such as trees~\cite{Guo2008IntervalGraphs,Dahlhaus1994Multiterminal}, and its computational hardness persists under constraints that commonly occur in practical applications, such as prohibiting terminal node removals. Although polynomial-time solutions exist for special classes such as interval~\cite{Guo2008IntervalGraphs} and co-bipartite graphs~\cite{PapadopoulosPermutation}, the general case resists tractable resolution.

Over the past decades, several approaches have been developed to cope with the problem's inherent complexity. These include  integer linear programming (ILP) models, constant-factor and logarithmic approximation algorithms~\cite{Garg1996Approx,Calinescu2003Multicuts}, and more recently, fixed-parameter tractable (FPT) strategies that exploit structural parameters of the input~\cite{Bousquet2011Multicut,Marx2011MulticutFPT}. However, classical algorithms often faces scaling challenges with the size and connectivity of the network, especially when the number of terminal pairs grows or the graph exhibits irregular or high-degree topology.  

In this work, we explore an alternative computational paradigm: quantum annealing, implemented on the D-Wave quantum processing unit (QPU). Quantum annealing is a metaheuristic that aims to find ground state-energy (i.e., optimal or near-optimal) configurations of binary optimization problems encoded as \textit{Quadratic Unconstrained Binary Optimization (QUBO)} models or their Ising equivalents~\cite{Glover2019QUBOTutorial,Farhi2000QAA}. D-Wave's architecture, designed specifically for such formulations, offers a unique opportunity to experimentally assess the performance of quantum hardware on NP-hard graph problems.

We focus specifically on tree-structured instances of the Vertex Minimum Multicut problem, for which QUBO formulations are constructed using path-based constraints. Our goal is not only to encode and solve these instances using quantum annealing, but also to evaluate the practical bottlenecks in the process. These include limitations in embedding due to hardware topology~\cite{CruzSantos2023DWaveQUBO}, chain stability~\cite{Bertuzzi2024Chains}, and the trade-offs involved in hybrid quantum-classical solvers such as those used by D-Wave's Advantage platform~\cite{DWaveDocs2024}.

By focusing on a well-established yet computationally challenging problem, we aim to evaluate the practical viability of quantum annealing for a variant that, to the best of our knowledge, has not been previously studied in this context. In addition to proposing a QUBO formulation, we analyze the impact of hardware constraints, embedding strategies, and solver behavior on both solution quality and runtime performance. The remainder of this paper is structured as follows. Section 2 formalizes the Vertex Minimum Multicut problem, discusses its complexity landscape, and outlines the path-based  formulation. Section 3 details the quantum annealing workflow and the role of D-Wave’s QPU. Section 4 presents empirical results and discusses practical observations. Section 5 concludes the paper.

\section{Problem Statement}

\textit{Multicut problems} are central in communication network design, reliability, and security, where one seeks to disconnect specific terminal pairs by removing a minimum set of graph elements (edges or vertices). Several variants of the problem exist depending on the type of cut (edge vs. vertex) and whether cuts can include terminal nodes.

In this paper, we focus on the Restricted Vertex Minimum Multicut (RVMMC) problem, in which the goal is to remove a minimal number of \emph{non-terminal vertices} such that all given terminal pairs are disconnected. For brevity, we refer to this problem as the Vertex Minimum Multicut (VMMC) throughout the remainder of the paper.

Let \( G = (V, E) \) be an undirected graph and \( H = \{(s_i, t_i) \mid i = 1, \dots, k\}, \quad \text{where } k \in \mathbb{N}, \text{ and } H \subseteq V \times V \)
 a set of terminal pairs. The objective is to find a subset \( C \subseteq V \) of non terminal vertices
such that for every \( (s_i, t_i) \in H \), all paths connecting \( s_i \) and \( t_i \) are disrupted by at least one vertex in \( C \).

This problem is NP-complete even on trees of bounded degree~\cite{CruzSantos2023DWaveQUBO}, and intractable on graph classes like cographs and split graphs. However, it admits polynomial-time solutions on interval graphs~\cite{Guo2008IntervalGraphs}, permutation graphs, and co-bipartite graphs~\cite{PapadopoulosPermutation}. Moreover, the parameterized version is fixed-parameter tractable with respect to the size of the cutset~\cite{Bousquet2011Multicut,Marx2011MulticutFPT}. Related problems have been studied under different formulations. The multi-terminal vertex separator was addressed using Branch-and-Price methods in~\cite{Magnouche2020MTVS}. QUBO models have been proposed for the edge multicut on trees using D-Wave~\cite{CruzSantos2023DWaveQUBO} and for the multiway cut~\cite{Heidari2022QUBO} with optimized formulations. Table~\ref{tab:cut-comparison} summarizes key distinctions across multicut variants.
\begin{table}[htbp]
\caption{Comparison of Cut-Based Graph Separation Problems.}
\label{tab:cut-comparison}
\centering
\begin{tabular}{|p{2.1cm}|p{1.3cm}|p{1.3cm}|p{2.5cm}|}
\hline
\textbf{Problem} & \textbf{Cut Type} & \textbf{Terminal Removal} & \textbf{Complexity (General Graph)} \\
\hline
Min-Cut max flow & Edge & Allowed & Polynomial-time \\
\hline
Multiway Cut & Edge/Vertex & Depends on model & NP-Complete for $|T| \geq 3$ \cite{Dahlhaus1994Multiterminal} \\
\hline
Edge Multicut & Edge & Allowed & NP-Complete for $|H| \geq 3$ \cite{CruzSantos2023DWaveQUBO} \\
\hline
Restricted Vertex Multicut & Vertex & Not Allowed & NP-Complete, even on trees \cite{Guo2008IntervalGraphs} \\
\hline
\end{tabular}
\end{table}

\vspace{6pt}

We now introduce a path-based Binary Linear Program (BLP) to model VMMC.
Each vertex \( v \in V \) is assigned a binary variable \( x_v \in \{0,1\} \), where \( x_v = 1 \) if the vertex is selected for removal. Let \( P_{s_i, t_i}\)  denotes  the set of all \(s_i-t_i\) paths, \( i = 1 , \dots, k \in \mathbb{N} \) . The objective is to minimize the total number of removed vertices while satisfying two constraints:

\begin{equation}
\label{eq:blp}
\begin{aligned}
\min \quad & \sum_{v \in V} x_v \\
\text{s.t.} \quad
& x_v = 0 \quad && \forall v \in V_H \quad \text{(C1)} \\
& \sum_{v \in \pi} x_v \geq 1 \quad && \forall (s_i, t_i) \in H, \forall \pi \in P_{s_i, t_i} \quad \text{(C2)} \\
& x_v\in \{0,1\}
\end{aligned}
\end{equation}

\vspace{4pt}

The constraint (C1) forbids the removal of terminal vertices. Constraint (C2) ensures that there exists no path between each terminal pair source target verticies. This formulation does not account for redundancy among paths, but provides a suitable base for QUBO conversion and quantum optimization experiments.

\section{Quantum Optimization}

Quantum computing introduces a novel paradigm for addressing combinatorial optimization problems by leveraging quantum mechanical phenomena such as superposition and tunneling, providing an efficient scheme to efficiently solve large-scale instances to explore large solution spaces~\cite{Abbass}. Although general NP-complete problems, including SAT, are widely believed to be not solvable in polynomial time by quantum computers (that is, they are not in BQP)~\cite{Abbass}, quantum approaches can still offer substantial advantages in practice. In particular, Grover’s algorithm provides at most a quadratic speedup in the unstructured search setting~\cite{grover1996fast}, reinforcing the theoretical limits of black-box optimization. However, these limitations do not rule out the possibility of significant quantum speedups on structured problem instances commonly encountered in real-world settings~\cite{aaronson2022structure}.

Motivated by this nuance, quantum heuristic methods such as quantum annealing are receiving increasing attention for solving NP-hard combinatorial problems where structure can be exploited. In this work, we focus on the application of quantum annealing to instances of the VMMC. By formulating the latter as a Quadratic Unconstrained Binary Optimization (QUBO) model, we explore the effectiveness of D-Wave’s quantum processing unit.

\subsection{Quantum Computing}\label{AA}

Optimization problems can be addressed using gate-based quantum algorithms, such as Grover’s adaptive search or the Quantum Approximate Optimization Algorithm (QAOA), or using analog approaches such as Quantum Annealing (QA)~\cite{Alex,Abbass}. For a quantum device to process an optimization problem, it must be formulated in a compatible mathematical model. This is typically a degree-two Boolean polynomial encoded as either a QUBO model or an Ising Hamiltonian for most of the analogue machines in contrast to higher degree Hamiltonian for digital machines. In gate-based quantum computers, the QUBO is mapped to quantum circuits consisting of unitary operations acting on qubits in a Hilbert space. The evolution of the quantum state leads to a probabilistic measurement, collapsing to a classical solution.

In contrast, quantum annealers like D-Wave's QPU physically realize optimization by embedding the QUBO or Ising model directly into hardware. The goal is to find the ground state—i.e., the lowest energy configuration—of the encoded cost function. The system is initialized in the ground state of an easy-to-prepare Hamiltonian and gradually evolved toward the problem Hamiltonian under a time-dependent Schrödinger equation~\cite{Farhi2000QAA}. If this evolution is sufficiently slow, the adiabatic theorem ensures that the system remains in its ground state throughout, yielding the optimal or near-optimal solution.

In both models, the quality of the final solution critically depends on the formulation of the problem, hardware constraints, and noise resilience, making the encoding strategy a central design component of quantum optimization workflows~\cite{venegas2018cross}.

\subsection{Quantum Annealing}\label{QA}

Quantum annealing (QA) is a metaheuristic to solve combinatorial optimization problems by exploiting quantum fluctuations, especially quantum tunneling, to escape local minima and explore low-energy configurations.  QA relies on a physical process governed by the \textit{quantum adiabatic theorem}~\cite{Farhi2000QAA}.

In this framework, the optimization problem is encoded into the ground state of a \textit{problem Hamiltonian} \( H_p \), acting on a Hilbert space \( \mathcal{H} = (\mathbb{C}^2)^{\otimes n} \) associated with \( n \) qubits. The system is initialized in the ground state of a \textit{driver Hamiltonian} \( H_0 \in \mathcal{L}(\mathcal{H}) \), where \( \mathcal{L}(\mathcal{H}) \) denotes the space of bounded linear operators on \( \mathcal{H} \). The full system evolves according to a time-dependent Hermitian operator:
\begin{equation}
    H(t) = A(t) H_0 + B(t) H_p, \quad t \in [0, T],
\end{equation}
where \( A(t), B(t) \in C^\infty([0,T], \mathbb{R}) \) are smooth, real-valued annealing schedules satisfying \( A(0) > 0, B(0) = 0 \) and \( A(T) = 0, B(T) > 0 \). Smoothness ensures differentiability at all orders, which is essential for the adiabatic condition.

If the system evolves slowly enough and the instantaneous spectral gap \( \Delta(t) = E_1(t) - E_0(t) \) between the ground state and the first excited state remains strictly positive for all \( t \in [0, T] \), the adiabatic theorem guarantees that the system remains in its instantaneous ground state throughout the evolution, ending in a state close to the ground state of \( H_p \). On D-Wave quantum annealers, \( H_p \) is implemented as an Ising Hamiltonian:
\begin{equation}
    H_{\text{p}} = \sum_{i=1}^{n} h_i Z_i + \sum_{1 \le i < j \le n} J_{ij} Z_i Z_j,
\end{equation}
where \( Z_i \) is the Pauli-\( Z \) operator acting on qubit \( i \), with eigenvalues \( \pm 1 \), and \( h_i, J_{ij} \in \mathbb{R} \) are tunable scalar parameters representing local fields and pairwise couplings, respectively. In the computational basis, the Hamiltonian is diagonal and defines a classical energy function:
\[
E(\mathbf{s}) = \sum_i h_i s_i + \sum_{i < j} J_{ij} s_i s_j, \quad \mathbf{s} \in \{-1, +1\}^n,
\]
whose minimization corresponds to identifying the ground state of \( H_p \).

\subsection{QUBO Formulation}

Let \( \mathbf{x} = (x_1, x_2, \ldots, x_n) \in \{0,1\}^n \) be a vector of binary decision variables. A general \textit{pseudo-Boolean function} is a multivariate polynomial \cite{CruzSantos2023DWaveQUBO} \( f : \{0,1\}^n \rightarrow \mathbb{R} \) defined as:

\begin{equation}
f(\mathbf{x}) = \sum_{S \subseteq \{1,\dots,n\}} c_S \prod_{j \in S} x_j,
\end{equation}

where \( c_S \in \mathbb{R} \). We define \( \deg(f) = \max \{ |S| : c_S \neq 0 \} \), when \( \deg(f) = 2 \), the function is called a \textit{quadratic pseudo-Boolean function}, which can be  written as:

\begin{equation}
f(\mathbf{x}) = \sum_{i=1}^{n} u_i x_i + \sum_{1 \leq i < j \leq n} w_{ij} x_i x_j.
\end{equation}

This yields the canonical QUBO form:
\begin{equation}
\min_{\mathbf{x} \in \{0,1\}^n} \mathbf{x}^\top Q \mathbf{x}, \quad Q \in \mathbb{R}^{n \times n},
\end{equation}
where \( Q \) is a symmetric cost matrix. In this model, the diagonal entries \( Q_{ii} \) represent linear terms and the off-diagonal entries \( Q_{ij} \) capture pairwise interactions.
The QUBO formulation is closely related to the Ising model in physics. Via a variable substitution \( x_i = \frac{1 + s_i}{2} \) where \( s_i \in \{-1, +1\} \), one can map the QUBO into an equivalent Ising Hamiltonian. This allows direct hardware implementation on quantum annealers such as D-Wave, which evolve the system toward the ground state of the encoded energy landscape.
In practice, the BLP is transformed into QUBO form by incorporating constraint violations as penalty terms. These penalty terms are weighted by hyperparameters that must be tuned to enforce the constraints while avoiding numerical instabilities or misrepresentations of the energy landscape~\cite{Glover2019QUBOTutorial}.

Let \( M_1, M_2\in \mathbb{R}\) denote the penalty coefficients for the terminal, and path constraints, respectively. Given binary variables \( x_v \in \{0,1\} \), the QUBO formulation becomes:

\begin{equation}
\label{eq:qubo1}
\begin{aligned}
\mathbf{x}^\top H_p\mathbf{x} = & \sum_{v \in V} x_v \\
& + M_1 \left( \sum_{v \in V_H} x_v \right)^2 \quad \text{(C1)} \\
& + M_2 \sum_{(s_i, t_i) \in H} \sum_{\pi \in P_{s_i,t_i}} \left( \sum_{v \in \pi} (1 - x_v) \right)^2 \quad \text{(C2)}
\end{aligned}
\end{equation}

This expression encodes the objective and constraints as penalty terms into a single quadratic function that the quantum annealer minimizes. While this formulation avoids slack variables, alternative versions may introduce auxiliary binary slack variables to transform inequality constraints into equalities~\cite{Glover2019QUBOTutorial}. For example, a path constraint of the form 
\( \sum_{v \in \pi} x_v \geq 1 \)
can be reformulated as 

\[
\left( \sum_{v \in \pi} x_v + \sum_{j=0}^{\lceil \log_2 M_{\pi} \rceil - 1} 2^j \cdot y_j^{\pi} - 1 \right)^2,
\forall \pi\in P_{s_i,t_i}\]
where \( y_j^{\pi} \in \{0,1\} \) are binary slack variables and \( M_{\pi} \) is the number of nodes in the path. This formulation increases expressiveness but also introduces additional variables and quadratic terms, which may impact embedding feasibility and resource usage on quantum hardware.

\subsection{D-Wave's Quantum Annealing Workflow}\label{AA}

Once the VMMC problem is formulated as a QUBO, solving it on D-Wave’s quantum annealer involves a nontrivial compilation pipeline. In particular, the path-based encoding of terminal separations introduces a formulation overhead, as the number of constraints scales with the set of enumerated paths. The resulting QUBO graph must then be embedded into D-Wave’s hardware topology—such as Pegasus or Zephyr—which imposes strict qubit connectivity constraints. Embedding logical variables onto chains of physical qubits is itself NP-hard and requires additional qubits and couplers~\cite{Bertuzzi2024Chains}. 

To address scalability and noise-related issues, we adopt D-Wave’s hybrid solver based on the racing strategy~\cite{DWaveDocs2024}. This approach concurrently evaluates multiple classical-quantum solver instances, dynamically prioritizing those with faster convergence.  decomposition enables better utilization of both classical heuristics and quantum tunneling for refining partial solutions.

\section{Numerical Analysis}

This section presents a performance evaluation of QUBO for the Vertex Minimum Multicut VMMC problem on tree-structured instances. The study investigates solver behavior, D-Wave hardware bottlenecks, and the effect of parameter tuning.
We generate random unweighted trees where each terminal pair shares a unique path. This guarantees feasibility of the path-based QUBO formulation while preserving nontrivial embedding complexity. Each instance is built as a random spanning tree with carefully selected non-adjacent terminal pairs to avoid degenerate cuts. The study is focused on 9 types of instances varying from number of terminals pair equals to 3 until 100 and number of verticies from 20 to 400

To ensure realistic performance, we focus on instances where the size and structure reflect practical limitations in embedding and qubit resources. This is particularly relevant when scaling up the number of terminal pairs.

\subsection{QUBO Parameter Tuning}

We conduct a grid search over penalty coefficients , using simulated annealing to estimate feasibility likelihoods and ground state convergence. Monte Carlo sampling is also applied to validate penalty robustness.
Penalty weights were calibrated empirically to avoid excessive dominance (which could flatten the energy landscape) or weakness (which risks infeasible solutions). Final values are scaled based on graph size and the number of terminal pairs.

\subsection{Solvers and Evaluation Metrics}

Performance is assessed using four key metrics. First, solution quality is evaluated through the \textit{optimality gap}, defined as \( \frac{f_{\text{found}} - f^*}{f^*} \times 100\% \), where \( f^* \) is the optimal cost obtained via the MILP solver when tractable. In larger cases, where MILP is computationally infeasible, only relative comparison across solvers is reported. Sampling time, which includes annealing and postprocessing duration, is measured to reflect the computational cost. Embedding difficulty is quantified using chain statistics (length, break count), while feasibility rate captures the proportion of solutions that satisfy all constraints imposed by the original BLP formulation

\begin{table}[h]
\centering
\caption{Solver Backends.}
\label{tab:solvers}
\begin{tabular}{|l|p{4.8cm}|}
\hline
\textbf{Type} & \textbf{Backend} \\
\hline
Quantum Annealing & D-Wave Advantage 4.1 QPU \\
Classical Heuristic & Simulated Annealing (custom) \\
Exact Solver & CPLEX \\
Hybrid Solver & Leap Hybrid Solver (Racing) \\
\hline
\end{tabular}
\end{table}


Additional metrics include QPU sampling time and solution feasibility. Simulated Annealing is tuned using inverse temperature ranges \((0.1, 10.0)\) and \((0.1, 20.0)\), sweep counts \{100, 200, 500\}, and both geometric and linear cooling schedules. Shot counts vary from 10 to 1000. For each configuration, we log the penalty weights \(M_1\), \(M_2\), the chosen annealing schedule, and sampling parameters. Chain statistics are only discussed in relation to feasibility breakdowns when relevant.

\subsection{D-Wave Backend Configuration}

Experiments were executed on D-Wave’s Advantage System 4.1 (eu-central-1), which supports the Pegasus P16 topology. Embedding was performed using the \textit{MinorMiner} algorithm with emphasis on minimizing chain lengths and avoiding chain breaks. The annealing time was fixed at 20$\mu$s, selected within the system’s available range.

The hybrid solver was run in \textit{racing mode}, where multiple classical and quantum samplers are launched in parallel. This improves response time and solution robustness. Coupling strength \( J \) was tuned over the extended range \([-2, 1]\) to stabilize qubit chains and reduce fragmentation.

\subsection{Findings and Discussion}

\begin{figure*}[htbp]
\centering
\includegraphics[width=0.71\linewidth]{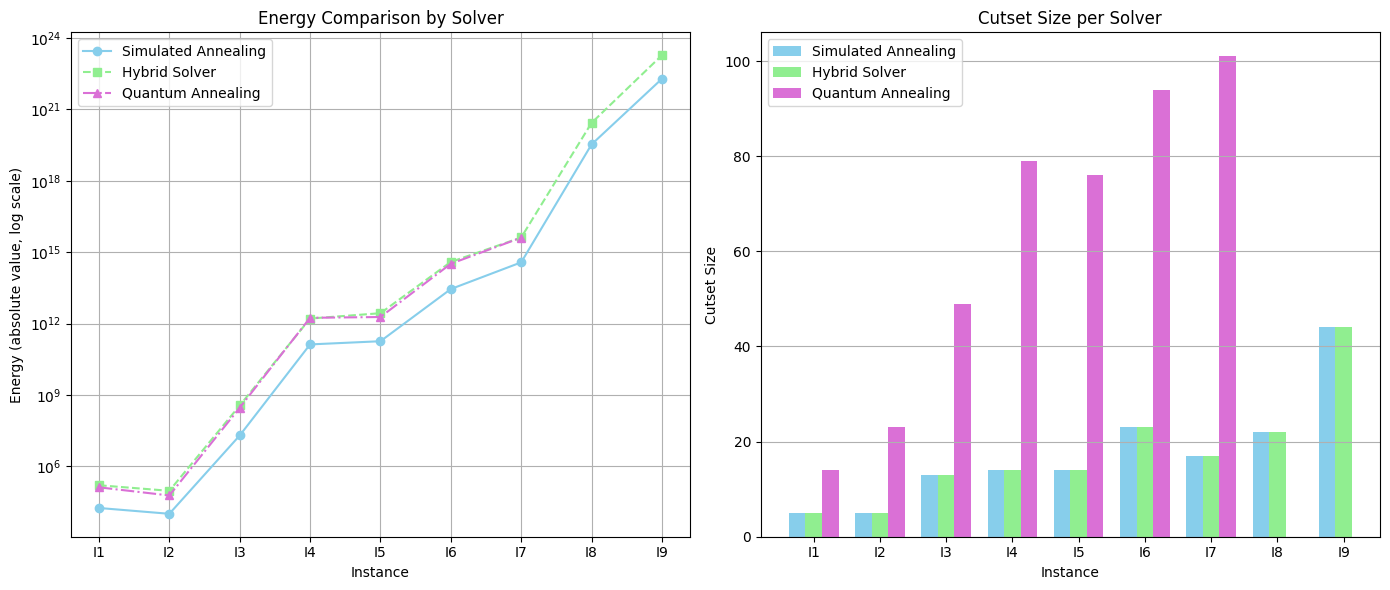}
\caption{Left: Solver energy comparison (absolute value log scale) across instances. Right: Associated cutset sizes indicating solution feasibility.}
\label{fig:solution-comparison}
\end{figure*}

\begin{figure}[htbp]
    \centering
    \includegraphics[width=1\linewidth]{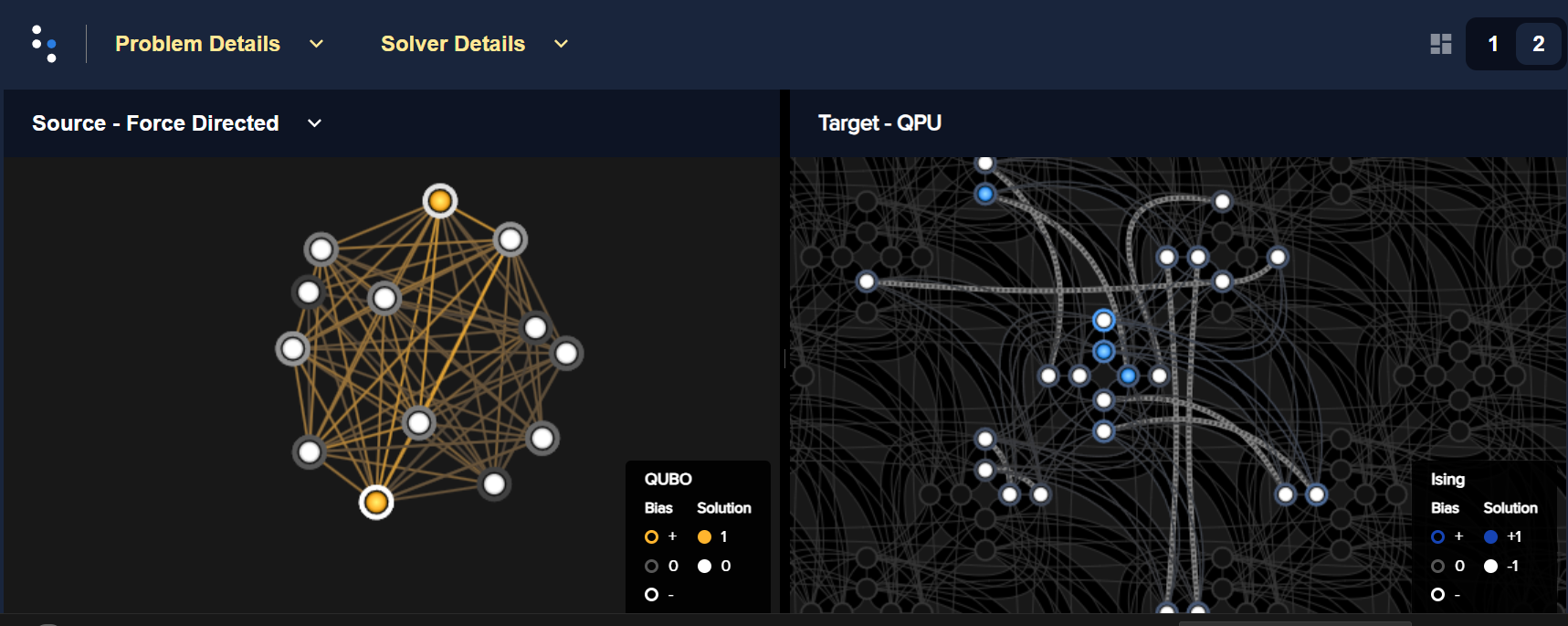}
    \caption{Source-to-QPU mapping and chain structure using D-wave inspector tool for a small instance.}
    \label{fig:embedding-chain}
\end{figure}

We evaluate nine instances of increasing complexity by simultaneously scaling the number of vertices and terminal pairs. Instance I1 is the smallest, with \(|H| = 3\) and \(|V| = 24\), while instance I9 is the largest, with \(|H| = 100\) and \(|V| = 450\).
As shown in Fig.~\ref{fig:solution-comparison}, we compare the minimum QUBO energy and the size of the associated cutset obtained from four solver types. Simulated annealing (SA) consistently achieves better cutset sizes, indicating better quality solutions—about 10 to 15\% gap—, although it does not always reach the lowest energy configurations. Hybrid solvers tend to minimize energy more effectively than classical methods, yet this advantage does not necessarily translate into better solutions due to limitations in the energy landscape encoding.

A key observation is that the quantum annealing solver often achieves low-energy solutions that are infeasible with respect to the original VMMC constraints. For example, in Instance I7, both the hybrid and quantum annealers reach comparable QUBO energies, but the quantum solver’s cutset deviates by over 300\% from the exact solution. Beginning with Instance I5, quantum annealing fails to yield any feasible solution, and for instances I8 and I9, embedding the QUBO into the QPU was not possible regardless of the embedding strategy. Multiple approaches were attempted, including MinorMiner, LazyFixedEmbeddingComposite, Uniform Torque Compensation, and clique-based embeddings, all without success.

In smaller instances, successful embeddings were possible due to lower overhead. However, as the number of terminal pairs increases, so does the number of path constraints and quadratic terms in the QUBO, leading to complex connectivity requirements. These exacerbate the embedding process and introduce chain instability, often preventing valid solutions from being recovered during unembedding.


To better understand degradation in quantum annealing performance, we analyzed variable embeddings (Fig.~\ref{fig:embedding-chain}). Improper chain strengths often result in broken chains, which introduce infeasibility even when energy values appear low. In the visualized instance, increasing chain strength improves embedding stability, enabling the recovery of feasible solutions. This highlights the importance of careful unembedding strategies and accurate majority-vote decoding in extracting valid variable assignments.

\begin{figure}[htbp]
\centering
\includegraphics[width=1\columnwidth]{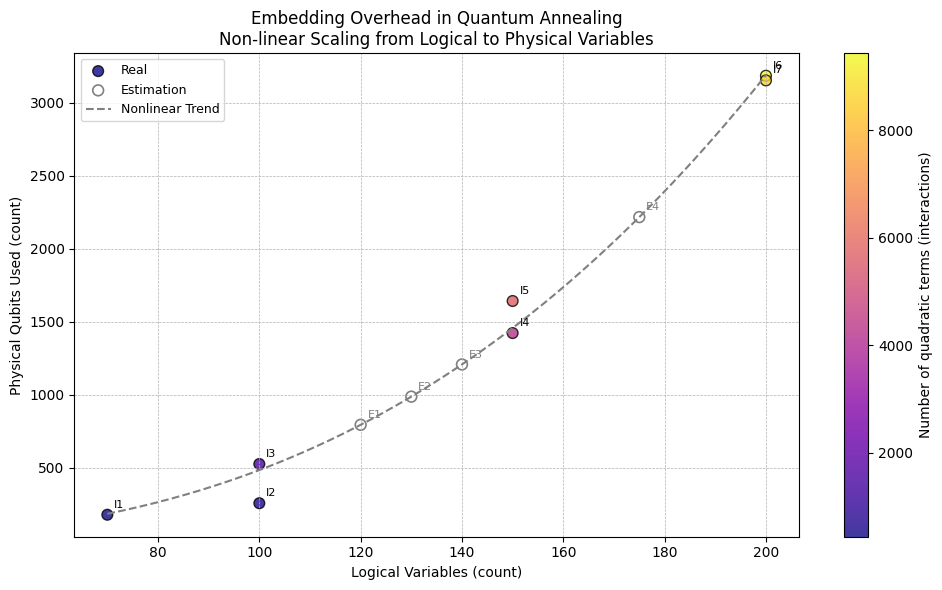}
\caption{Embedding overhead: non-linear scaling from logical to physical qubits in D-Wave quantum annealing.}
\label{fig:embedding-overhead}
\end{figure}

The non-linear relationship between logical variables and physical qubits is further illustrated in Fig.~\ref{fig:embedding-overhead}. Even instances with the same number of logical variables exhibit drastically different qubit counts, depending on the density of interaction terms. This is intrinsic to how QUBOs encode pairwise constraints and reflects the limitations of the Pegasus topology, which supports an average chain length of about 5 under ideal connectivity.

\begin{figure*}[htbp]
\centering
\includegraphics[width=0.80\linewidth]{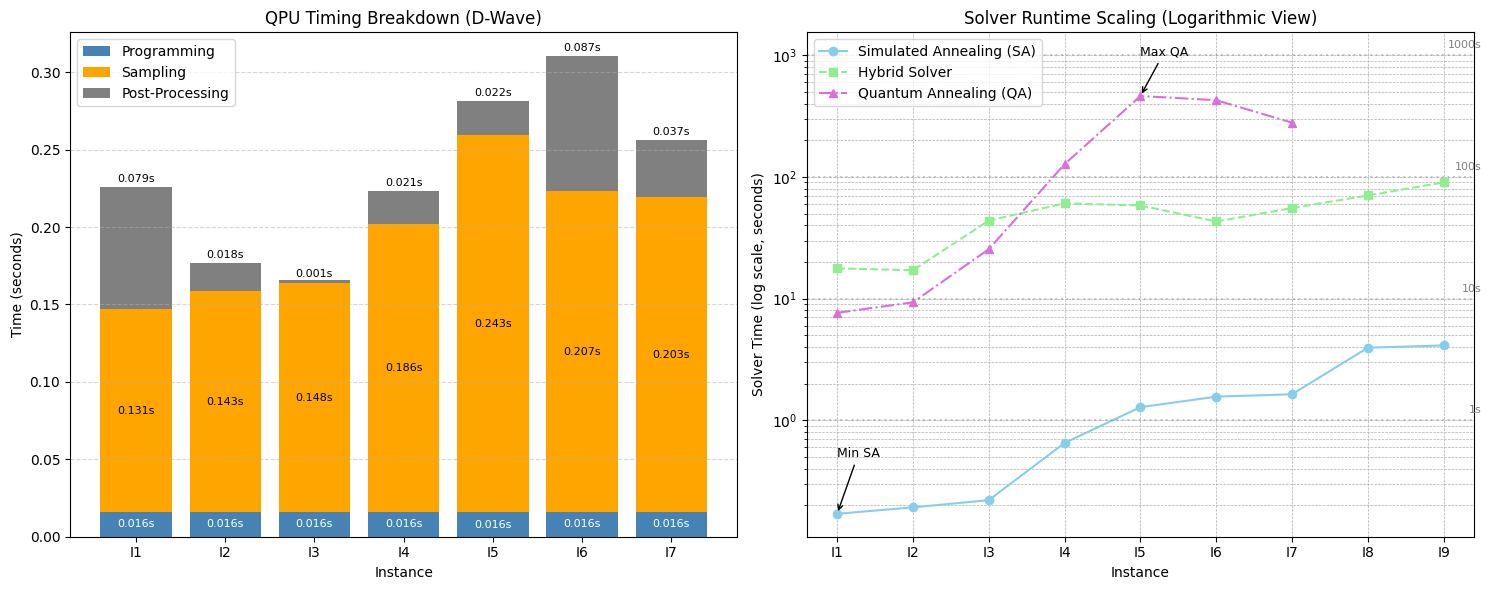}
\caption{Solver runtime comparison and QPU timing breakdown across instances.}
\label{fig:solver-timing}
\end{figure*}

Finally, Fig.~\ref{fig:solver-timing} compares solver runtimes. While the QPU’s physical annealing time remains low—dominated by sub-millisecond sampling—the total runtime is significantly affected by the compilation pipeline, including QUBO instantiation, embedding, and unembedding. Simulated annealing provides the best trade-off between runtime and solution quality for small to medium instances. Quantum annealing achieves competitive runtimes for small instances but quickly becomes impractical due to embedding failures. The hybrid solver balances both regimes by decomposing QUBOs into tractable subproblems, making it a compelling choice for intermediate instance sizes.


\section{Conclusion and future work}

This work presents an experimental evaluation of quantum annealing for the Restricted Vertex Minimum Multicut (VMMC) problem, modeled as a QUBO and solved using D-Wave’s quantum hardware. While the problem is naturally suited to combinatorial optimization, encoding it as a path-based QUBO introduces significant overhead as the number of terminal pairs increases.

Although quantum annealing often reaches low-energy configurations, these do not consistently yield feasible solutions due to broken qubit chains and embedding limitations. The embedding process, which maps logical variables to physical hardware qubits, introduces non-linear overhead stemming from hardware connectivity constraints and QUBO density. As a result, scalability suffers and solution quality degrades in larger instances. Hybrid quantum-classical solvers demonstrate greater robustness by decomposing the problem into tractable substructures, achieving feasible solutions at larger scales. Simulated annealing also remains competitive in both runtime and cutset quality.

Future work will focus on leveraging problem structure to reduce the number of quadratic interactions in the QUBO, enabling more compact embeddings. More strategic control over chain strength and adaptive embedding heuristics may improve solution feasibility in quantum-only runs.

While full quantum advantage remains elusive, our findings show that hybrid workflows and tailored formulations can effectively contribute to real-world cybersecurity optimization.

\end{document}